\documentclass[lettersize,journal]{IEEEtran}
\usepackage{amsmath,amsfonts}
\usepackage{algorithmic}
\usepackage{algorithm}
\usepackage{array}
\usepackage[caption=false,font=normalsize,labelfont=sf,textfont=sf]{subfig}
\usepackage{textcomp}
\usepackage{stfloats}
\usepackage{url}
\usepackage{verbatim}
\usepackage{graphicx}
\usepackage{cite}

\usepackage[nolist,nohyperlinks]{acronym}
\usepackage{todonotes}

\begin{acronym}
\acro{rfsoc}[RFSoC]{Radio-Frequency System-on-Chip}
\acro{daq}[DAQ]{Data Acquisition}
\acro{soc}[SoC]{System-on-Chip}
\acro{mkid}[MKID]{Microwave Kinetic Inductance Detector}
\acro{mmc}[MMC]{Magnetic Microcalorimeter}
\acro{squid}[SQUID]{Superconducting Quantum Interference Device}
\acro{pl}[PL]{Programmable Logic}
\acro{ps}[PS]{Processing System}
\acro{dac}[DAC]{Digital-Analog Converter}
\acro{adc}[ADC]{Analog-Digital Converter}
\acro{vna}[VNA]{Vector Network Analyzer}
\acro{iir}[IIR]{Infinite Impulse Response}
\acro{qic}[QIC]{Quantum Interface Controller}
\acro{sdr}[SDR]{Software-defined radio}
\acro{rfdc}[RFdc]{RF data converter}
\acro{cirque}[cirque]{Communication interface to readout electronics for quantum experiments}
\acro{fir}[FIR]{Finite Impulse Response}
\acro{axi}[AXI]{Advanced eXtensible Interface}
\acro{csv}[CSV]{Comma-Separated Values}
\acro{psd}[PSD]{Power Spectral Density}
\end{acronym}

\begin{document}

\title{Real-Time Readout System Design for the BULLKID-DM Experiment: Enhancing Dark Matter Search Capabilities}

\author{T.~Muscheid, R.~Gartmann, L.~E.~Ardila-Perez, A.~Acevedo-Rentería, L.~Bandiera, M.~Calvo, M.~Cappelli, R.~Caravita, F.~Carillo, U.~Chowdhury, D.~Crovo, A.~Cruciani, A.~D’Addabbo, M.~De Lucia, G.~Del Castello, M.~del Gallo Roccagiovine, D.~Delicato, F.~Ferraro, M.~Folcarelli, S.~Fu, M.~Grassi, V.~Guidi, D.~Helis, T.~Lari, L.~Malagutti, A.~Mazzolari, A.~Monfardini, D.~Nicolò, F.~Paolucci, D.~Pasciuto, L.~Pesce, V.~Pettinacci, C.~Puglia, D.~Quaranta, C.~M.~A.~Roda, S.~Roddaro, M.~Romagnoni, G.~Signorelli, F.~Simon, M.~Tamisari, A.~Tartari, E.~Vázquez-Jáuregui, M.~Vignati, K.~Zhao
        % <-this % stops a space
%\thanks{This paper was produced by the IEEE Publication Technology Group. They are in Piscataway, NJ.}% <-this % stops a space
\thanks{
Manuscript received April 19, 2021; revised August 16, 2021.

Please see the Acknowledgment section of this article for the author
affiliations.

}}

% The paper headers
\markboth{Journal of \LaTeX\ Class Files,~Vol.~14, No.~8, August~2021}%
{Shell \MakeLowercase{\textit{et al.}}: A Sample Article Using IEEEtran.cls for IEEE Journals}

\IEEEpubid{0000--0000/00\$00.00~\copyright~2021 IEEE}
% Remember, if you use this you must call \IEEEpubidadjcol in the second
% column for its text to clear the IEEEpubid mark.

\maketitle

\begin{abstract}
The BULLKID-DM experiment aims to detect WIMP-like potential Dark Matter particles with masses below 1\,GeV/c². Sensing these particles is challenging, as it requires nuclear recoil detectors characterized by high exposure and an energy threshold in the order of 100\,eV, thus exceeding the capabilities of conventional semiconductor detectors. BULLKID-DM intends to tackle this challenge by using cryogenic \acp{mkid} with exceptional energy thresholds to sense a target with a total mass of 800\,g across 16 wafers, divided into over 2000 individually instrumented silicon dice.
The \acp{mkid} on each wafer are coupled to a single transmission line and read using a frequency division multiplexing approach by the room-temperature data acquisition.

In this contribution, we describe and assess the design of the room-temperature readout electronics system, including the selected hardware components and the FPGA firmware which contains the real-time signal processing stages for tone generation, frequency demultiplexing, and event triggering. 
We evaluate the system on the ZCU216 board, a commercial evaluation card built around a \ac{rfsoc} with integrated high-speed DACs and ADCs, and connected it to a custom-designed analog front-end for signal conditioning. 
%The system has been characterized and verified by coupling it to the demonstrator detector consisting of 3 wafers with 60 instrumented silicon dice each.
\end{abstract}


\section{Introduction}
\IEEEPARstart{O}{ne} of the main objectives of modern physics is the experimental proof of the existence of dark matter. Several theoretical models exist, one of which proposes Dark Matter to be WIMP-like particles \cite{Bertone_2018}. Although many combinations of cross-sections and energies of these WIMP-like particles have already been excluded in previous experiments, they remain a highly promising candidate \cite{Billard_2022}. Current research is eager to search for particles with low energies ($\leq 1\,keV$) and small cross sections ($\leq 10^{-40} cm^2$), which are still in the blind region of the dark matter map, above the neutrino floor. Detecting such low-energetic hypothetical particles is tremendously difficult, as they rarely interact with matter and therefore cannot be measured with conventional sensors. A promising method to reach the required sensitivity is cryogenic crystal detection, which relies on measuring lattice vibration (phonons) generated during the interaction of dark matter particles with the target mass. With BULLKID-DM, we aim to search for dark matter with cryogenic sensors featuring an energy threshold around 200\,eV and an unprecedented number of detectors. The operation of these cryogenic devices is complex and requires special \ac{daq} systems with low-noise hardware components and custom firmware for online data processing. In this contribution, we aim to give a detailed picture of the custom \ac{daq} system and its usability for operating frequency-multiplexed low-temperature detectors. A brief introduction to the BULLKID-DM experimental setup is given in chapter \ref{sec_bullkid} to derive the specific requirements for the room-temperature readout electronics. In chapter \ref{sec_daq_architecture}, the architectural details of the hardware and firmware of the system are described. Subsequently, the calibration routine for preparing and performing a measurement is described and characterized in chapter \ref{sec_calibration} before the noise contribution of the readout system is evaluated in chapter \ref{sec_system_characterization}. 

\section{BULLKID-DM Experiment}
\label{sec_bullkid}
\IEEEpubidadjcol The BULLKID-DM experiment employs cryogenic sensors to detect nuclear recoil events. In this setup, \acfp{mkid} are used to measure the phonons that are produced when Dark Matter particles interact with a target, e.g. silicon or germanium. The experiment employs 2300 dice of the target material with a mass of 0.35\,g each, totaling a mass of 800\,g, split across 16 wafers. Each 4-inch wafer is cut into 145 squares, and each die is coupled to a \ac{mkid} with resonance frequencies ranging from 800\,MHz to 1\,GHz \cite{Delicato_2024}. This defines the requirements for the room-temperature readout electronics: 16 input and output interfaces to the cryostat, one pair for each wafer. On each of these lines, a frequency comb containing the resonance frequencies needs to be generated by the DAQ system to stimulate the resonators. To reconstruct the raw detector signals, the modulated frequency combs are digitized, frequency demultiplexed, and downconverted. 
Based on the design choices of the BULLKID-DM experiment, several requirements for the room-temperature readout electronics can be derived, which are summarized in table \ref{tab_reqs}. 

\begin{table}[htbp]
	\caption{Requirements for the room-temperature readout electronics.}
	\begin{center}
		\begin{tabular}{|l|c|}
			\hline
			\textbf{Parameter} & \textbf{Parameter value} \\
			\hline
			Number of cryogenic coaxial lines & 16  \\
			Frequency range of resonators & 800\,MHz - 1\,GHz  \\
			Frequency resolution of stimulus & 100\,Hz \\
			Number of detectors per readout line & 145  \\
			Total number of detectors & 2300 \\
			Detector bandwidth & 100\,kHz \\
			Tone power & $<$ -5\,dBm \\
			Amplitude and phase noise & 100\,dBc @ 0.1-10\,kHz \\
			\hline
		\end{tabular}
		\label{tab_reqs}
	\end{center}
\end{table}

For preliminary evaluation of the system performance, a detector demonstrator consisting of three 3-inch wafers with 60 dies each has been constructed. While the final experiment with the full-scale detector will be located at the Gran Sasso underground laboratory in Italy for a minimized background activity, this demonstrator is operated in a lab at Sapienza university in Rome for testing the individual parts of the system.

\section{DAQ System Architecture}
\label{sec_daq_architecture}
The DAQ system for the BULLKID-DM experiment is part of the \ac{qic} family, a series of data acquisition systems that leverage the software-defined radio concept and the configurable real-time firmware modules. Building on the existing readout system for the ECHo-100k experiment \cite{Muscheid_2024}, the framework has been adapted for BULLKID by modernizing the hardware to utilize \acf{rfsoc} devices and tailoring the \ac{pl} firmware to its new requirements. This highlights the broad applicability of the QIC system across different detector types, including MMCs, MMBs, and MKIDs.

\subsection{Hardware}
\label{subsec_hardware}
\begin{figure}[htbp]
	\centering
	\includegraphics[width=0.9\columnwidth]{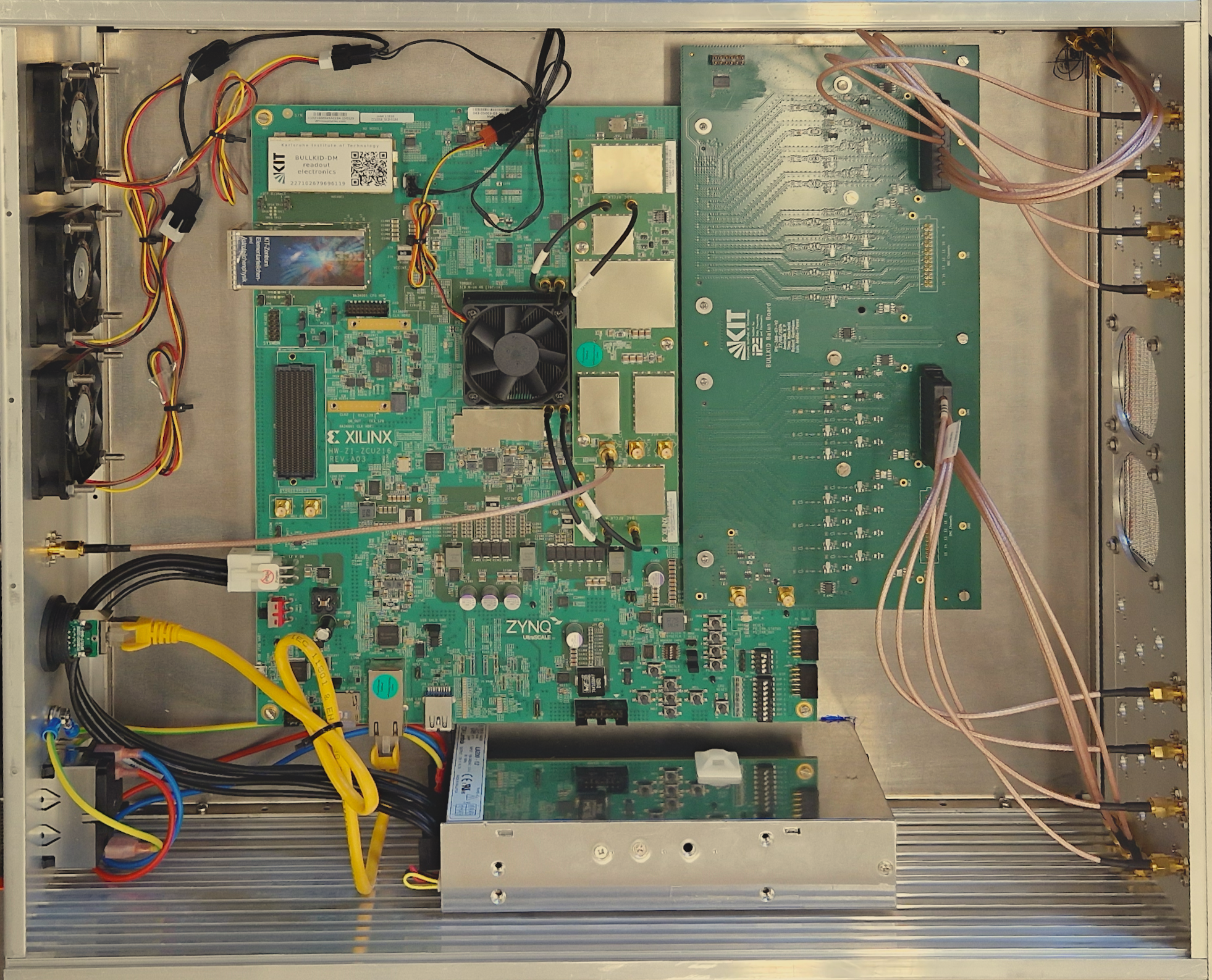}
	\caption{BULLKID-DM DAQ system setup. The 19-inch rack-mounted enclosure houses the ZCU216 evaluation board with the custom front-end, along with the power supply unit, cooling fans, cabling, and accessories.}
	\label{fig_daq_system}
\end{figure}

The core of the DAQ system is the AMD-Xilinx XCZU49DR \ac{rfsoc} device, which contains 16 \acp{dac} operating up to 9.85\,GSPS and 16 \acp{adc} up to 2.5\,GSPS directly integrated into the silicon. Since this device belongs to the Ultrascale+ family, migrating the firmware and software stack from the previously used ZynqUS+ ZCU19EG chip on our custom DTS100G board \cite{Muscheid_2023} is straightforward. AMD-Xilinx additionally offers the ZCU216 evaluation board built around this device. The evaluation board comes with the CLK104 mezzanine card, which comprises a 10\,MHz oscillator as a reference and several RF synthesizers to generate the high-frequency clocks for the analog converters. The clock tree of the entire system is derived from this source to ensure synchronization. For optimized phase stability, the system can alternatively be locked to a 10\,MHz input reference clock. For analog signal conditioning, we designed a custom front-end board that attaches to the RFMC-2.0 connectors of the ZCU216 board. It contains Balanced-Unbalanced (BalUn) transformers, amplifiers, and two stages of low-pass aliasing filters for each readout line, both in the TX and RX directions. The FMC+ connector is currently unused but will integrate the energy calibration system, which is under active development. The ZCU216 and its mezzanine boards are installed in a 19-inch housing for convenient rack mounting next to the cryostat, as shown in figure \ref{fig_daq_system}.

\subsection{Firmware}
\begin{figure}[htbp]
	\centering
	\includegraphics[width=\columnwidth]{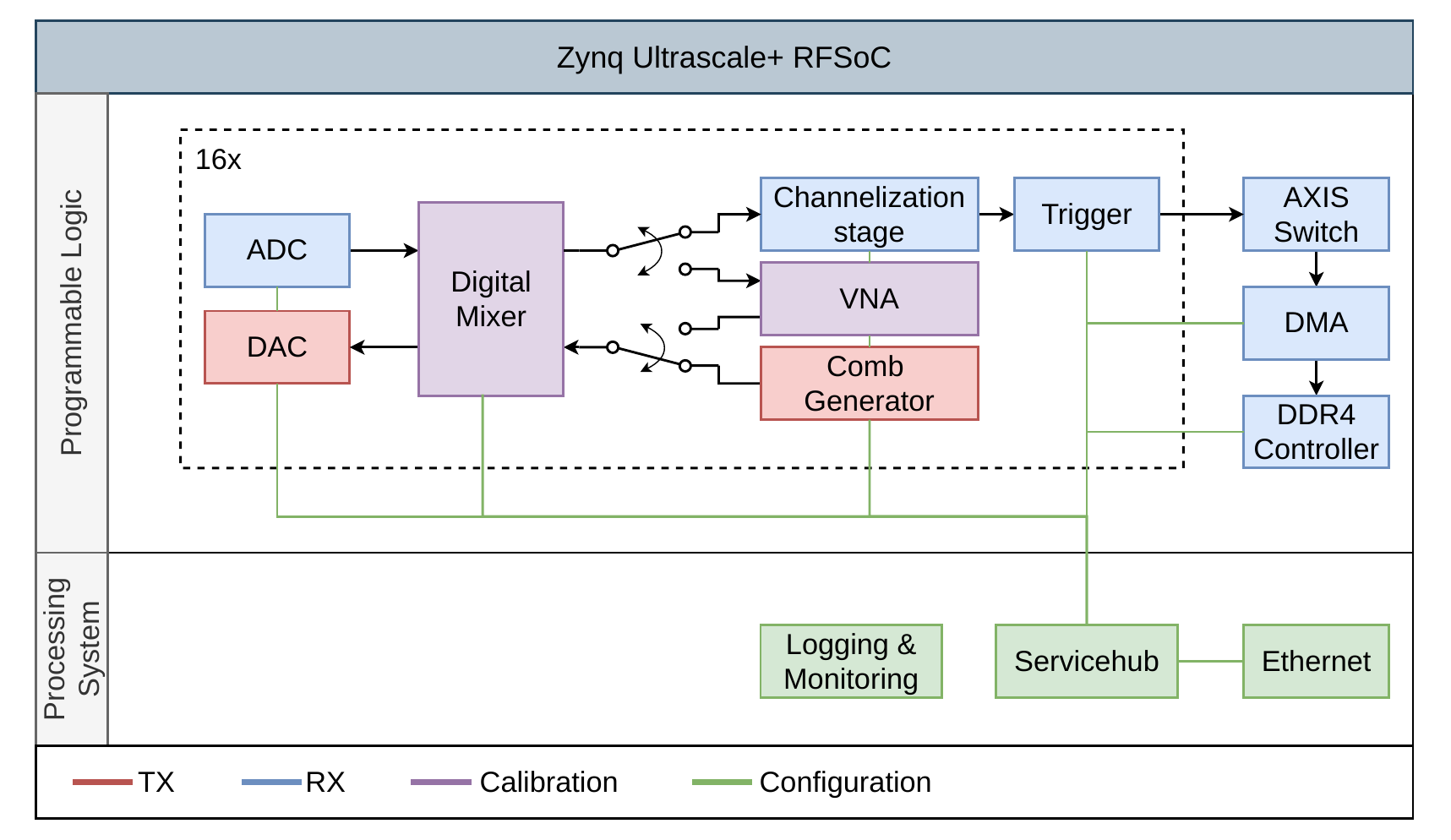}
	\caption{Block diagram of the firmware implemented on the \ac{rfsoc}. The upper section illustrates the real-time processing chain running on the \ac{pl}, while the logging and the custom Servicehub software for user interaction are implemented on the \ac{ps} part of the \ac{soc}.}
	\label{fig_firmware}
\end{figure}

The \ac{daq} system is used to perform the first stage of data processing with the goal of reducing the data rate for subsequent offline data analysis. Following the \ac{sdr} approach, the generation of the stimulus, demodulation of the frequency comb, and downconversion are all performed in the digital domain. A high-level overview of the firmware is shown in figure \ref{fig_firmware}. The implementation details of the individual modules in the \ac{pl} are presented in section \ref{sec_firmware_details}. Some of the modules depicted in figure \ref{fig_firmware} are not needed for data acquisition but for initial system calibration, which is described in section \ref{sec_calibration}.  

\subsection{Build system}
Creating a firmware image for the \ac{rfsoc} devices containing both a Linux operating system as well as the real-time signal processing of the FPGA is a complex task. Due to the many dependencies between the \ac{pl} and the \ac{ps}, a mature framework is required. While we previously used the open-source framework Yocto as our build system, we recently developed a custom toolset called SoCks \cite{Fuchs_2025} optimized for our use case of building firmware for the AMD-Xilinx \acp{soc}. Compared to Yocto, SoCks minimizes the dependencies between the individual firmware blocks, thereby improving maintainability and simplifying upgrades to newer kernel versions. The online processing chain on the \ac{pl} is developed using VHDL to maximize implementation efficiency and provide precise control over the low-level circuitry. The VHDL block diagram is created using both AMD-Xilinx IP cores and custom modules.

\subsection{User interaction}
\label{subsec_user_control}
The entry point for communication between experimentalists and the DAQ system is a custom program called Servicehub, running in the \ac{ps} of the \ac{rfsoc} \cite{Karcher_2021}. The Servicehub comprises a gRPC server that receives commands from remote machines and forwards the information to the shared memory of \ac{ps} and \ac{pl}, where the FPGA modules interact via \ac{axi}. One advantage of gRPC is its support for a multitude of programming languages, for which the appropriate API, so-called stubs, can be autogenerated from the message protobuf declaration. We created a Python package called \ac{cirque} \cite{Muscheid_2025} as the API to our system that can be installed on the user's PC via pip. This package contains convenient wrappers for all commands our system provides. These commands allow the experimentalists to configure the firmware modules during runtime. For instance, users can update the individual tones of the frequency comb or change the trigger parameters. %\ac{cirque} is intended to be the low-level interface to the system without any additional features. For higher-level functionality and more complex commands, we developed a second Python package that encapsulates \ac{cirque}. As part of our user control library, we offer functionality to automatically configure the DAQ system and to parse the binary data and convert it into HDF5 data format for subsequent post-processing.
The low level interface of \ac{cirque} is abstracted by another Python-based wrapper for automated system configuration and HDF5 file conversion of the acquired measurement data.

\section{Firmware details}
\label{sec_firmware_details}
As depicted in figure \ref{fig_firmware}, the real-time signal processing consists of several stages. Here, the individual modules implemented in the \ac{pl} are examined in detail.

\subsection{Converter configuration}
\label{sec_coverter_config}
On the \ac{rfsoc}, the converters (see section \ref{subsec_hardware}) are configured using the AMD-Xilinx \ac{rfdc} IP core. In this application, the \acp{dac} are configured to operate at 5\,GSps, while the \acp{adc} operate at their maximum sampling frequency of 2.5\,GSps. The RF synthesizers on the CLK104 mezzanine card are configured to generate the required clocks to drive the \acp{dac} and \acp{adc}, respectively. Internally, the \ac{rfdc} is configured to derive a 125\,MHz clock from the \ac{dac} reference clock, which is used by the firmware modules in the \ac{pl}. Operating the converters at higher sampling frequencies is beneficial, as it moves the spurs to higher Nyquist zones further away from the readout frequencies \cite{Gartmann_2024}. This simplifies the attenuation of image frequencies with low-pass filters.

\begin{figure}[htbp]
	\centering
	\includegraphics[width=\columnwidth]{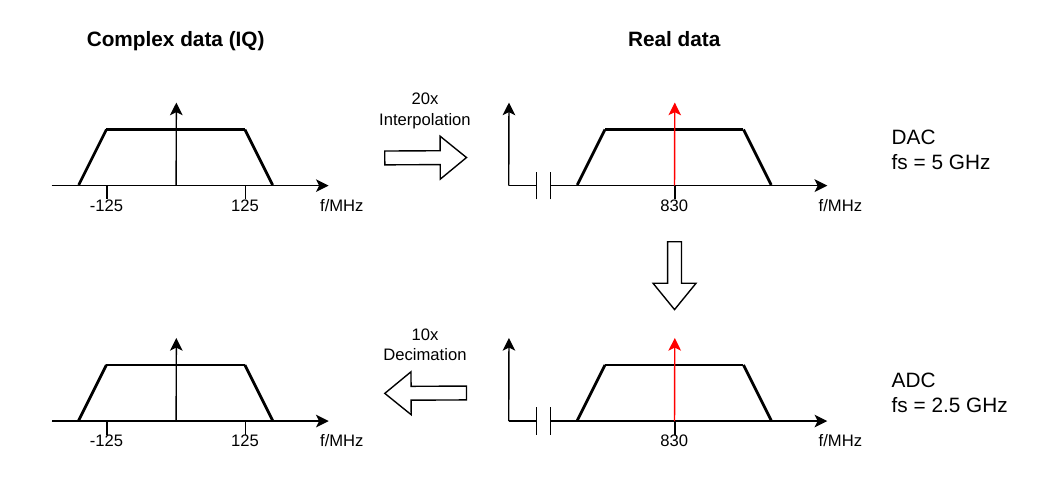}
	\caption{Configuration of the converters within the \ac{rfdc} IP core of the \ac{rfsoc}. All 16 converter channels are configured equally and share a common input reference clock.}
	\label{fig_rfdc}
\end{figure}

As listed in table \ref{tab_reqs}, the resonators require an instantaneous bandwidth of 250\,MHz. By setting the interpolation factor in the \acp{dac} to 20 and the decimation factor in the \acp{adc} to 10, the bidirectional data rate in the digital domain is set to 250\,MHz. Additionally, the internal mixer within the \ac{rfdc} is enabled on both the TX and RX sides and set to 830\,MHz to convert the digital signal between baseband and the target frequency band. The configuration of the \ac{rfdc} is depicted in figure \ref{fig_rfdc}.

\subsection{Stimulus generation}
\label{subsec_stimulation}
To read out frequency-multiplexed cryogenic detectors, the resonators have to be stimulated by generating a frequency close to their resonance. Superimposing many of these stimulation tones leads to a so-called frequency comb. The DAQ system presented here is able to synthesize such a comb where the individual frequencies, as well as their amplitude and phase, can be chosen at run-time.
Since the BULLKID-DM detector does not require the stimulation tone frequency to be changed dynamically, a simple approach for generating the tones is implemented: After defining the frequency, amplitude, and phase of each tone in the comb, the time domain samples of each tone are summed up on the user PC and sent to the \ac{daq} system, which stores them in a memory. The module in the FPGA replays the content of this memory cyclically. To ensure a continuous waveform without phase jumps at the end of the cycle, only frequencies for which an integer number of periods fits into the reserved memory are allowed. This results in the frequency resolution being dependent on the memory size $n$, as shown in equation \eqref{eq_stimulus_res}.

\begin{equation}
	 \Delta f = \frac{f_{s}}{n}\label{eq_stimulus_res}
\end{equation}

With a target frequency resolution $\Delta f =100\,Hz$ and a sampling rate $f_{s} = 250\,MHz$, a minimum number of $n=2,500,000$ samples need to be stored. Given that each sample is stored as 2 times 16-bit complex data, a memory size of 80\,Mb is required for each cryogenic line. Since the \ac{rfsoc} offers 60.5\,Mb of on-chip storage, this memory can only be used for a prototype system with reduced frequency resolution. The samples for the high-resolution stimulus signal will instead be stored on the DDR, occupying 1.28\,Gb (160\,MB) of memory and 16\,GB/s throughput. Both memory consumption and throughput demands are within the specification of the ZCU216 board. As an alternative, more resource-efficient approach, frequency comb generation based on digital signal processing without precalculated samples is currently under investigation.

\subsection{Channelization stage}
\begin{figure}[htbp]
	\centering
	\includegraphics[width=\columnwidth]{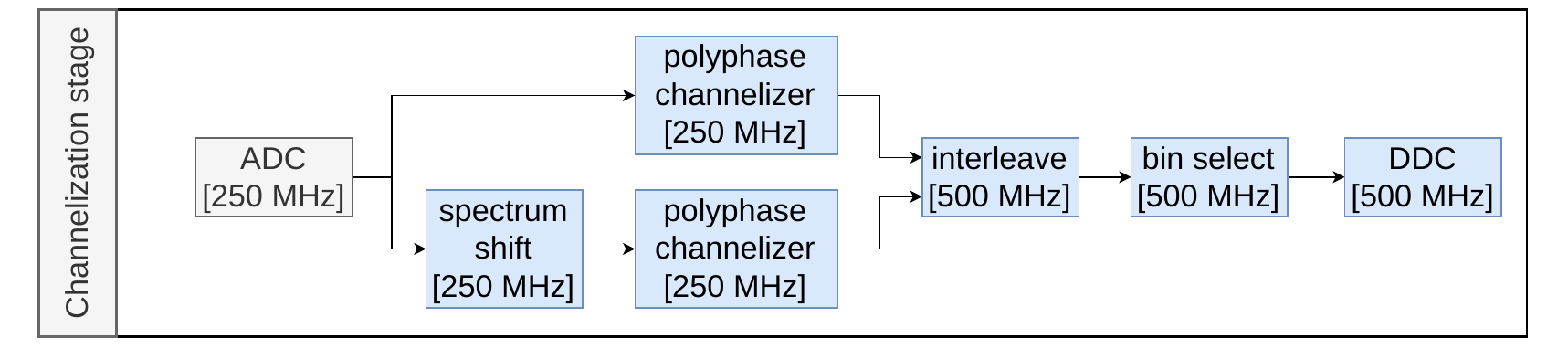}
	\caption{Block diagram of the full digital channelization stage. It converts the wideband input signal from the \ac{adc} containing the modulated frequency comb into a time-division multiplexed data stream containing the reconstructed and separated detector signals.}
	\label{fig_channelization}
\end{figure}

The initial step of the digital signal processing is to separate the modulated carrier tones into channels and to reconstruct the individual detector signals. This channelization process is divided into several steps, the first of which occurs directly within the \ac{adc}, where the frequency band containing the carrier signals is mixed down to baseband (see \ref{sec_coverter_config}. In the digital domain, several steps are required to reconstruct the raw detector signals, as depicted in figure \ref{fig_channelization}. 

First, a polyphase channelizer separates the input signal with sampling rate $f_{s}=250\,MHz$ into 64 equidistant sub-bands with 1.9\,MHz filtered bandwidth each \cite{Xilinx_2013}. However, the critically sampled filter bank has blind intervals, where the readout tones are fully attenuated. Therefore, we implement another polyphase channelizer operating on a frequency-shifted copy of the input data in parallel. Its frequency shift is 1.95\,MHz, hence the second channelizer fully covers the blind intervals of the first instance. The two sets of 64 channels are combined into a single data stream of 500\,MHz by interleaving. Subsequently, the data is processed by a channel-select module, which discards empty channels and duplicates others if more than one carrier tone is present. Finally, a second stage for downmixing contains a variable mixer to demodulate the carrier tones. It is equipped with an \ac{fir} filter module that limits the signal bandwidth to 100\,kHz, as required according to table \ref{tab_reqs}. Additionally, the \ac{fir} filter performs a sample rate decimation by a factor of 20, yielding a sampling rate $f_{det} = 195\,312.5\,kSps$ per resonator, as calculated in equation \eqref{eq_det_f}. At the end of this channelization stage, the raw detector signals are successfully reconstructed and can be further processed. 

\begin{equation}
	f_{det} = \frac{f_{s}}{n}*\frac{1}{N} = \frac{250\,MHz}{64}*\frac{1}{20} = 195\,312.5\,kHz
	\label{eq_det_f}
\end{equation}

For readout of the demonstrator system with 60 resonators on 3-inch wafers, this architecture with a total number of 128 channels is sufficient. For the final experiment setup with 145 resonators on each 4-inch wafer, the channelization stage will be extended by a third parallel polyphase channelizer with 64 sub-bands to support a total number of 192 channels. This extension can easily be implemented, since we designed the VHDL modules to be configurable using a set of generics. All modules can be optimized according to the specific needs of the application, making our framework flexible and versatile.

\subsection{Online Trigger}
\label{subsec_trigger}
After reconstruction of the raw detector signals, various trigger algorithms can be applied to detect and extract pulses, thereby greatly reducing the data rate. These algorithms include a simple digital differentiation that compares two successive samples, a moving average filter, and an \ac{iir} filter. Once the trigger detects a pulse, a data package of up to 1024 samples is stored, which equals to a timespan of around 5.25\,ms. The data package includes pre-trigger samples and metadata such as the detector index, timestamp, and the type of trigger used. 

To improve resource utilization, the raw detector data streams of all 16 cryogenic lines are combined and processed by a single trigger module. Since it is clocked at 500\,MHz and designed for pipelined operation, it can process up to 2560 channels with the sampling rate calculated in equation \eqref{eq_det_f}. Given that 2300 detectors need to be read out in the full-scale scenario (see table \ref{tab_reqs}) this approach is appropriate for zero deadtime operation.
This solution also simplifies synchronization of the trigger in the third dimension: with all detectors processed by a single module, detecting particles interacting with the target mass on different layers of the detectors is possible directly on the FPGA, enabling directional nuclear recoil analysis. 

\subsection{Data storage}
\label{subsec_data_storage}
The channelized and triggered data is subsequently stored in DDR4 memory. From the DDR, the data stream can either be stored on the platform itself or directly streamed to the user PC via Ethernet for post processing and analysis. Additionally to the triggered data packages including metadata, the firmware allows acquisition of the raw data at the output of the channelization stage. For data taking, two distinct methods are available: to acquire a specific number of samples or perform a measurement for a specific time span a ``snapshot'' can be executed. For long-term data acquisition, continuous measurements can be started and will run until explicitly stopped by the user. Independent of the measurement method, the data will be stored in binary form to keep the overhead of the data packets as small as possible.

\subsection{Firmware scalability}
\begin{figure}[htbp]
	\centering
	\includegraphics[width=\columnwidth]{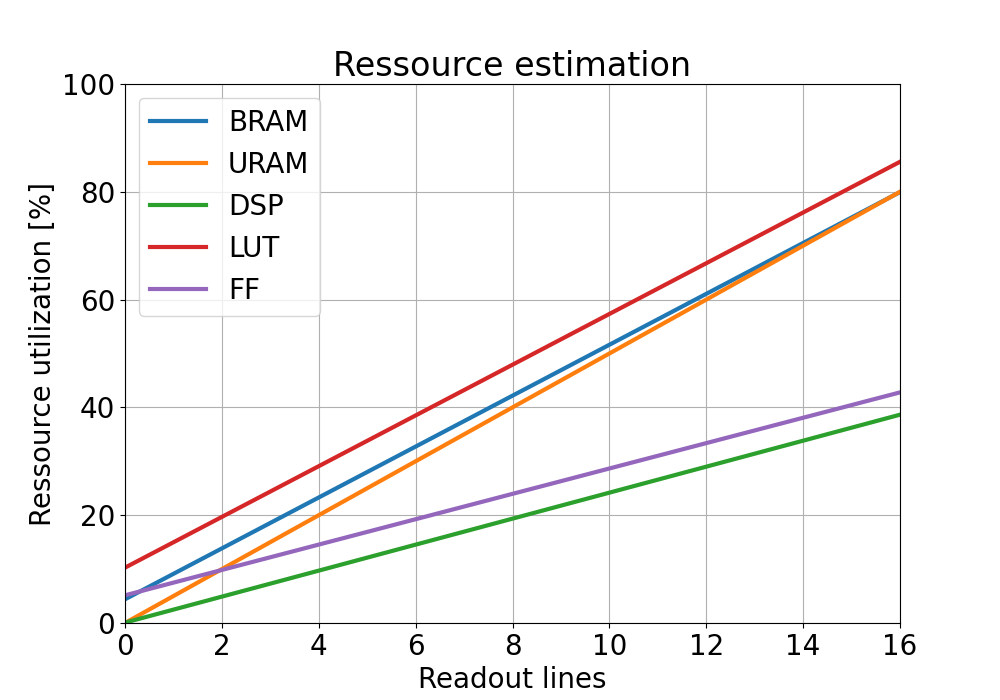}
	\caption{Projected resource estimation on the \ac{pl} for up to 16 processing chains. The offset in LUTs and BRAM resources results from the implementation of the storage module, whose size is independent of the number of readout chains.}
	\label{fig_ressource_estimation}
\end{figure}

One important aspect of the full-scale readout electronics is the scalability of the firmware. For the current demonstrator setup, the goal is to optimize a single chain, while the experiment requires integrating all 16 processing chains into the \ac{pl}. This scalability requirement was considered early in the design process and has been periodically verified. As shown in figure \ref{fig_ressource_estimation}, the available resources on the XCZU49DR device are sufficient to implement the full-scale design with up to 16 processing chains. Given the headroom in DSP slices, more complex trigger algorithms for improved sensitivity may be implemented in the future. Additionally, unused memory may enable further improvement of the stimulus frequency resolution.

\section{System calibration}
\label{sec_calibration}
Before starting a measurement, several calibration steps need to be performed to maximize the system's performance. The main firmware modules described above are therefore extended to support these calibration steps without the need for additional measurement equipment.

\subsection{Resonator detection}
While the resonators on the multiplexer have been designed to specific frequencies, the actual resonator frequencies depend on many parameters and vary with every cooldown cycle of the cryostat. To properly configure the stimulus generation module described in section \ref{subsec_stimulation}, these frequencies need to be determined. Previously, this was done using a \ac{vna}, requiring manual rewiring of the cryogenic lines. It is more convenient to directly integrate the \ac{vna} functionality into the firmware of the DAQ system, offering standalone operation. Additionally, this allows self-configuration of the device: after scanning the frequency range of interest, the resonator frequencies are determined automatically, and the values are written into a \ac{csv} file, which is used for precalculating the time-domain samples of the frequency comb.
 
\subsection{Resonator characterization}
\begin{figure}
	\centering
	%\subfloat{\includegraphics[width=0.5\columnwidth]{figures/circle_prerotation.pdf}}
	%\subfloat{\includegraphics[width=0.5\columnwidth]{figures/circle_postrotation.pdf}}
    \subfloat{\includegraphics[height=0.45\columnwidth]{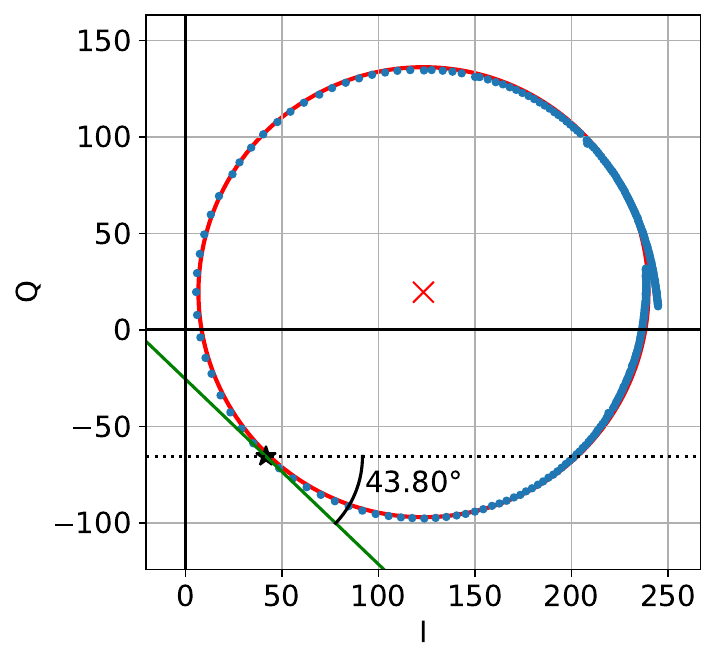}}
    \hfill
	\subfloat{\includegraphics[height=0.45\columnwidth]{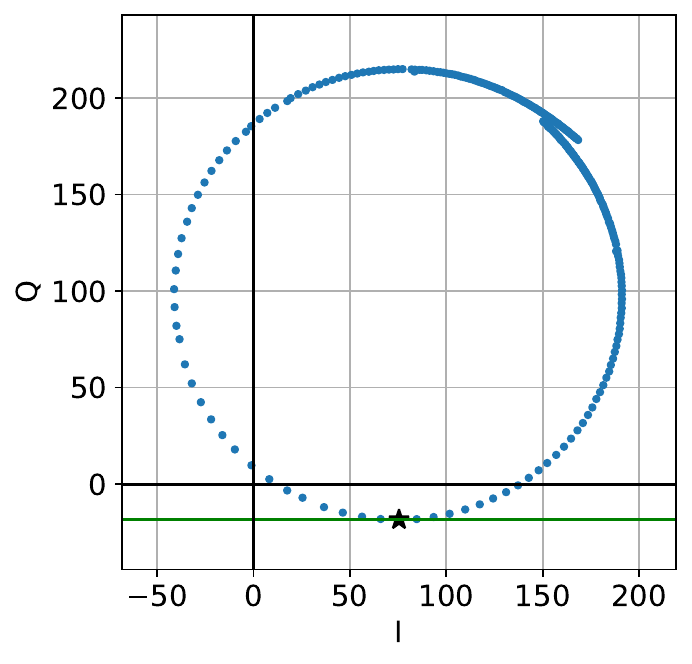}}
	\caption{Graphical representation of the phase rotation. A circle is fitted to the resonator data, and the angle of the tangent at the actual operation point is calculated (left). This angle is subsequently used to rotate the circle, resulting in the tangent having phase 0 (right). Notice the different position of the circles in the IQ plane.}
	\label{fig_circle_fit}
\end{figure}

Using the previously mentioned \ac{vna} feature of our system gives a rough estimate of the resonance frequency. However, to reach the maximum sensitivity, the optimal readout frequency and power for each resonators needs to be determined. The characterization of these parameters is carried out by sweeping the readout frequencies around the actual resonance frequency and measuring the $S_{21}$ parameter. By performing this frequency sweep with different tone powers, a two-dimensional dataset is created that can be used to estimate the optimum readout parameters.
In theory, this characterization can be performed with commercial measurement equipment by taking sweeps of various output powers with a \ac{vna}, as in the initial resonator detection. However, sweeping across all resonators on the multiplexer sequentially will take a long time, especially when a high frequency resolution is desired. A faster approach is supported by the platform: an additional mixer module is implemented that mixes the input signal with a user-defined frequency. Using the stimulation module with the generated stimulus tones as an input, simultaneous sweeps of all resonators can be achieved. The same mixer module is used to down-mix the frequency comb again so that no reconfiguration is required for the processing chain. Thus, the calibration for all resonators on a multiplexer can be obtained simultaneously, saving a lot of time. Plotting the acquired values for the measured frequencies in an I/Q diagram, shows that the phase of the $S_{21}$ parameter follows a circle when swept across the resonator (compare figure \ref{fig_circle_fit} left).
 
\subsection{Phase rotation} 
To detect pulses, the trigger algorithms described in section \ref{subsec_trigger} use either the phase or the amplitude of the complex signal as input, which requires analysis of both the I and the Q component of the signal. By rotating the resonance circle, we can make use of the small angle approximation and simplify the implementation of the trigger algorithms to only consider one of the two components. This rotation does not result in data loss, as both I and Q components of the signal are still stored for post processing. The angle of the rotation is also saved so that the original signal can be reconstructed if needed.
Usually, such a phase rotation operation requires an additional computation step, e.g. by using the CORDIC algorithm. However, with our electronics the rotation can be performed directly as part of the down-conversion chain. The NCO for the mixer has an input signal for the phase offset that can be used to delay the mixing frequency relative to the input signal. This delay results in a rotation of the down-mixed signal. The value for the phase offset is proportional to the desired angle of rotation and is determined using the resonance circle and a Python script.

\subsection{Energy calibration}
The final calibration stage is not required for the operation of the DAQ system, but for evaluation of the measurement results. To calculate a spectrum of the acquired data, the detectors need to be energy-calibrated. This energy calibration is achieved by an external system called LANTERN \cite{DelCastello_2024}, which sends a known amount of photons via optical fibers to the detectors. By measuring the resulting pulse height in the detector signal, each detector pixel can be calibrated individually. Current efforts focus on the integration of LANTERN into the DAQ system to automatically synchronize the photon firing and the data acquisition.

% This could also be included into the section where the channelization is described
\section{System Characterization}
\label{sec_system_characterization}
%Currently, the system is used in combination with the prototype detector system consisting of three wafers with 60 \acp{mkid} each. This demonstrator setup is used for characterizing the full-scale system in terms of noise performance. As a requirement for the experiment, the noise level for both amplitude and phase needs to be below 100\,dBc at 1\,kHz. The performance of the DAQ system is evaluated in loopback mode, with the \ac{dac} directly connected to the \ac{adc}, bypassing the cryogenic environment. For proper signal conditioning, a 20\,dB attenuator is added in the analog signal path to avoid exceeding the maximum input power of the \ac{adc}, given that the front-end board contains two 12\,dB amplifiers. As a first test, a single tone was generated by the system and downconverted by the processing chain after the \ac{adc}, leaving only a complex noise trace. The two components of the noise were subsequently used to calculate its amplitude and phase.

\begin{figure}[htbp]
	\centering
	\includegraphics[width=\columnwidth]{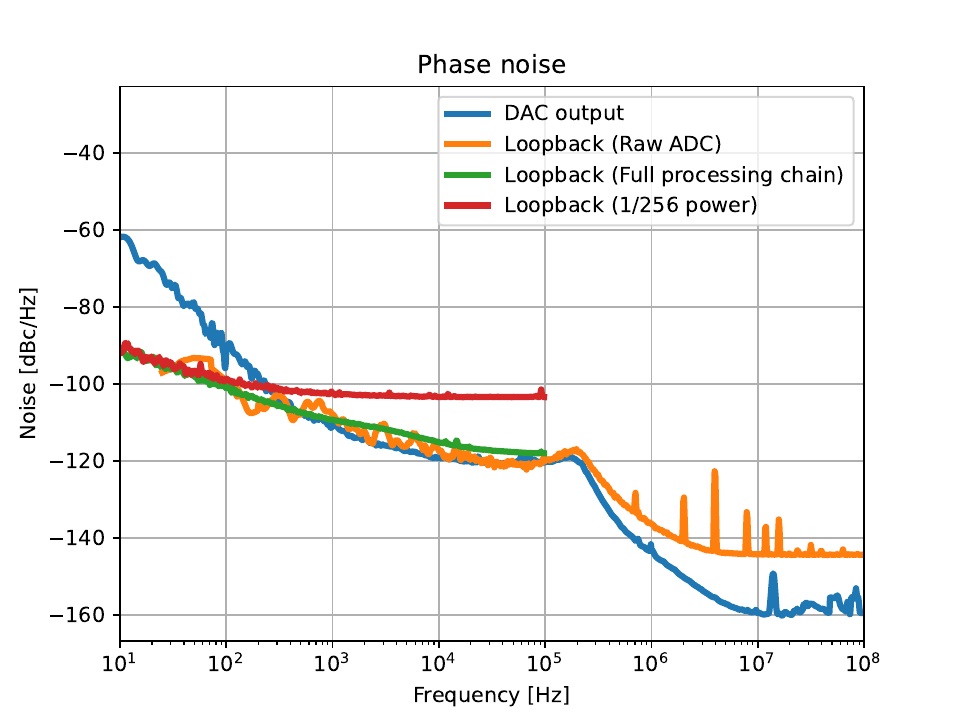}
	\caption{Phase noise analysis of the system using a single tone at 831\,MHz. In the low-frequency range relative to the carrier, the \ac{dac} (blue trace) is limiting performance, while the noise floor of the \ac{adc} (orange trace) is the limiting factor for the high-frequency range. Due to the sample rate reduction in the signal processing, the phase noise of the channelized signal (green, red trace) is limited to below 100\,kHz. The red trace showcases the phase noise of the system with a reduced tone power, simulating the full-scale operation with all readout tones sharing the available \ac{dac} power.}
	\label{fig_phase noise}
\end{figure}

Achieving the desired energy resolution requires a low-noise \ac{daq} system to prevent degradation of the detector sensitivity. To evaluate the performance of the processing chain and localize the major noise sources, the phase noise was measured at different stages in the processing chain. The phase noise of the \ac{dac} was evaluated using a Rohde\&Schwarz FSWP. For evaluation of the \ac{adc} and the signal processing chain, the \ac{daq} was operated in loopback-mode with the \ac{dac} output directly fed back into the \ac{adc}, thereby bypassing the cryogenic environment. The data was acquired using the data storage module introduced in \ref{subsec_data_storage} and the \ac{psd} was calculated using the python package pylpsd. We were able to identify the \ac{dac} as the main contributor to the system phase noise, further investigations showed that the root cause was a sub-optimal configuration of the clock chips on the CLK104 \cite{ad_2025}. Interestingly, tweaking the parameters of the RF synthesizers showed phase noise improvements also in loopback mode, with both the \ac{dac} and the \ac{adc} clocks derived from the same reference. Figure \ref{fig_phase noise} shows the system phase noise with improved configuration, achieving the required phase noise at the level of 100\,dBc at 100\,Hz - 10\,kHz. However, the \ac{dac} clock is still the limiting factor for frequency offsets below 100\,kHz, which are most relevant for the filtered and downconverted data. Further improvements can be achieved using high-quality synthesizers, reducing the spot noise at 1\,kHz up to 15\,dB compared to the optimized CLK104 configuration. Reducing the tone power to meet the requirement of simultaneously generating 145 tones increases the effect of quantization noise. Nevertheless, the goal of keeping the phase noise below 100\,dBc is fulfilled under these conditions, proving the readout system described here suitable for next-generation Dark Matter search with enhanced sensitivity.

\section{Conclusion}
The DAQ system described in this work showcases the flexibility and universality of the \ac{qic} \ac{sdr} framework we have developed over the years. The firmware modules were reconfigured to target the requirements of a new experiment and detector type. The modularity of the framework enabled the expansion of certain firmware modules to address the needs of direct Dark Matter search applications.
As shown in this contribution, the DAQ system fulfills the requirements for deployment in the full-scale BULLKID-DM experiment. It enables scalability to simultaneously process more than 2000 detectors on 16 cryogenic lines.
Characterization of the DAQ system performance shows excellent results and indicate a successful path towards the operation of the experiment at the Gran Sasso underground laboratory in Italy.

\section*{Acknowledgments}
We acknowledge the support of the Doctoral School \emph{“Karlsruhe School of Elementary and Astroparticle Physics: Science and Technology”}, the SECIHTI Project No. CBF-2025-I-1589 and DGAPA UNAM Grants No. PAPIIT IN105923 and IN102326. This work was further supported by the INFN, Sapienza University of Rome and co-funded by the European Union (ERC, DANAE, 101087663). Views and opinions expressed are however those of the author(s) only and do not necessarily reflect those of the European Union or the European Research Council. Neither the European Union nor the granting authority can be held responsible for them. We thank A. Girardi and M. Iannone of the INFN Sezione di Roma for technical support.

\subsection*{\textbf{Author's Affiliations}}
L.~E.~Ardila-Perez, D.~Crovo, R.~Gartmann, T.~Muscheid, and F.~Simon are with the Institute for Data Processing and Electronics at the Karlsruhe Institute of Technology, Hermann-von-Helmholtz-Platz 1 76344, Eggenstein-Leopoldshafen, Germany.

M.~Cappelli, M.~del Gallo Roccagiovine, M.~Folcarelli, L.~Pesce, D.~Quaranta, and M.~Vignati are with the Dipartimento di Fisica, Sapienza Università di Roma, P. le A. Moro 2, 00185 Roma, Italy and the INFN Sezione di Roma, P.le A. Moro 2, 00185 Roma, Italy.

D.~Delicato is with the Dipartimento di Fisica, Sapienza Università di Roma, P. le A. Moro 2, 00185 Roma, Italy, the INFN Sezione di Roma, P.le A. Moro 2, 00185 Roma, Italy, and the University Grenoble Alpes, CNRS, Grenoble INP, Institut Néel, 38000 Grenoble, France.

A.~Cruciani, G.~Del Castello, D.~Pasciuto, and V.~Pettinacci are with the INFN Sezione di Roma, P.le A. Moro 2, 00185 Roma, Italy.

M.~De Lucia, T.~Lari, D.~Nicolò, F.~Paolucci, C.~M.~A.~Roda, S.~Roddaro, and G.~Signorelli are with the Dipartimento di Fisica "Enrico Fermi", Università di Pisa, Largo Bruno Pontecorvo 3, 56127 Pisa, Italy and the INFN Sezione di Pisa, Largo Bruno Pontecorvo 3, 56127 Pisa, Italy.

F.~Carillo, M.~Grassi, C.~Puglia, and A.~Tartari are with the INFN Sezione di Pisa, Largo Bruno Pontecorvo 3, 56127 Pisa, Italy.

A.~Acevedo-Rentería and E.~Vázquez-Jáuregui are with the Instituto de Física, Universidad Nacional Autónoma de México, A.P. 20-364, Ciudad de México 01000, México.

L.~Bandiera, L.~Malagutti, and M.~Romagnoni are with the INFN Sezione di Ferrara, Via Saragat 1, 44122 Ferrara, Italy.

V.~Guidi and A.~Mazzolari are with the Dipartimento di Fisica e Scienze della Terra, Università di Ferrara, Via Saragat 1, 44100 Ferrara, Italy and the INFN Sezione di Ferrara, Via Saragat 1, 44122 Ferrara, Italy.

M.~Tamisari is with the Dipartimento di Neuroscienze e Riabilitazione, Università di Ferrara, Via Luigi Borsari 46, 44121 Ferrara, Italy and the INFN Sezione di Ferrara, Via Saragat 1, 44122 Ferrara, Italy.

M.~Calvo, U.~Chowdhury, and A.~Monfardini are with the University Grenoble Alpes, CNRS, Grenoble INP, Institut Néel, 38000 Grenoble, France.

A.~D’Addabbo, F.~Ferraro, S.~Fu, and D.~Helis are with the INFN Laboratori Nazionali del Gran Sasso, 67100 Assergi (AQ), Italy.

K.~Zhao is with the Gran Sasso Science Institute, Viale F. Crispi, 7 67100 L'Aquila and the INFN Laboratori Nazionali del Gran Sasso, 67100 Assergi (AQ), Italy.

R.~Caravita is with the INFN - TIFPA, Via Sommarive 14, 38123 Povo (Trento), Italy.

\end{document}